\documentstyle[12pt,psfig]{article}
\newcommand{\NP}[1]{ Nucl.\ Phys.\ {\bf #1}}
\newcommand{\ZP}[1]{ Z.\ Phys.\ {\bf #1}}
\newcommand{\RMP}[1]{ Rev.\ of Mod.\ Phys.\ {\bf #1}}
\newcommand{\PL}[1]{ Phys.\ Lett.\ {\bf #1}}

\newcommand{\PR}[1]{ Phys.\ Rev.\ {\bf #1}}
\newcommand{\PRL}[1]{ Phys.\ Rev.\ Lett.\ {\bf #1}}

\newcommand{\Od}{{\cal O}}

\newcommand{\Opd}{{\cal O}(p^2)}
\newcommand{\Opc}{{\cal O}(p^4)}

\newcommand{\tr}{\mbox{tr}}

\newcommand{\be}{\begin{equation}}
\newcommand{\ee}{\end{equation}}
\newcommand{\bea}{\begin{eqnarray}}
\newcommand{\eea}{\end{eqnarray}}

\newcommand{\Rea}{{\rm Re}\,}
\newcommand{\nn}{\nonumber}
\newcommand{\qq}{\langle\bar{q}q\rangle}
\newcommand{\qqo}{\langle0\vert\bar{q}q\vert0\rangle}
\newcommand{\mh}{\hat{m}}
\begin{document}

\hfill
\vbox{
      \hbox{SLAC-PUB-7865}
      \hbox{hep-ph/9806539}
      \hbox{June 1998}
      }

\vskip 1cm

\begin{center}
{\Large \bf The Hadronic Gas Chiral Phase Transition} 

\vskip .5cm

{\Large \bf within }

\vskip .5cm

{\Large \bf Generalized Chiral Perturbation Theory$^1$}

\vskip 1.5cm

J. R. Pel\'aez$^{2}$

\vskip .5cm

{\em Stanford Linear Accelerator Center}\\
{\em  Stanford University, 
Stanford, California 94309}

\end{center}

\footnotetext[1]{Research supported
by the Department of Energy under contract 
DE-AC03-76SF00515}
\footnotetext[2]{On leave of Absence from Departamento 
de F\'{\i}sica Te\'orica.\\
Universidad Complutense. 28040 Madrid. Spain.\\
electronic address:pelaez@slac.stanford.edu}
\begin{abstract}
We study the temperature evolution
of the $\langle \bar q q\rangle$
condensate below the chiral phase transition. 
The hadronic gas is described
using a virial expansion  within 
Generalized Chiral Perturbation Theory.
In such way, we can implement both the 
large or small chiral condensate scenarios and 
analyze the condensate dependence on the values of
the lightest quark masses.
\end{abstract}
{\footnotesize \hspace*{1cm} PACS:11.30.Rd, 12.39.Fe, 
21.65.+f, 51.30.+i,24.10.Cn,24.10.Pa}

\newpage

\section{Introduction}

The properties of QCD at finite temperature have raised
considerable interest in the literature (see \cite{review} 
and references therein). 
At low temperatures it seems that color is confined and chiral 
symmetry is spontaneously broken. However, from asymptotic freedom,
it is expected that at high temperatures, color will be liberated and
chiral symmetry restored. It is a matter of intense debate whether there 
should be one or two phase transitions, at what temperatures they 
would occur, and what their nature would be. 

Within the standard wisdom the quark chiral condensate $\qqo$ plays a 
central role in this problem, 
since it is assumed that chiral symmetry breaking is
produced by a strong condensation of quark-antiquark pairs \cite{GOR}.
However, in recent years, this hypothesis has been questioned, opening
the possibility of small, even vanishing, $\qqo$ scenarios 
\cite{GChPT1,GChPT2,Stern}. (Note that
we use $\qqo$ for the condensate at $T=0$ and $\qq$ in general).

The evolution of the quark condensate with the temperature has indeed 
been addressed using several approaches. In general, the properties
of $\qq$ can be derived from a somewhat idealized dilute pion gas,
which is commonly described using an effective Lagrangian formalism 
\cite{GeLe}, as we will do here, 
or by  means of finite temperature QCD
sum rules \cite{QCDSR}. In general, all these and other approaches 
\cite{other}
yield a rather consistent picture, although they usually have
the large condensate assumption built in.

In this work we want to know how the actual value of $\qqo$, as well as 
the light quark masses, can modify the behavior of
the chiral condensate, as for instance, with changes in the phase
transition temperature.
With that purpose, we will describe the pion gas by means of
the virial expansion and using the
interactions obtained from the
Chiral Perturbation Theory (ChPT) formalism \cite{GaLe}, 
although allowing for
a wide range of $\qqo$ values. Such a framework is usually referred
to as Generalized Chiral Perturbation Theory (GChPT) \cite{GChPT1,GChPT2}.
It should be noticed that we will be dealing with two effects:
First, at $T=0$,
 the size of $\qqo$, 
which may be different from the standard large value. 
Second, the evolution at finite 
temperature which is also changed through the modifications
in the meson interactions due to the different scheme of the explicit
chiral symmetry breaking. Our purpose
is to study what the interplay is of these two effects.

The plan of the paper is as follows: In section one we describe
briefly the ChPT and GChPT formalisms, focusing on
the relation between the quark and meson masses 
with the quark condensate. The next 
section is devoted
to the virial expansion for the pion gas, where we introduce
the temperature dependence. In section three we make the 
study of the condensate dependence
both on the temperature and the ratio of light quark masses,
using the $\Opc$ amplitudes of GChPT.
Next, in section four, we
estimate the contributions from heavier states, and in the conclusions
we summarize our results.

\section{Standard and Generalized Chiral Perturbation Theory}

When considering just three massless quark flavors,
the QCD Lagrangian
exhibits an $SU(3)_L\times SU(3)_R$ symmetry which,
even neglecting particle masses,
is not present in the physical spectrum. Instead,
we observe an approximate $SU_{L+R}(3)$ symmetry, which means
that  the $SU(3)_{L-R}$ chiral group has to be
 spontaneously broken. According to the Goldstone Theorem, there should be
eight massless Goldstone Bosons (GB), which are identified with the
pions, kaons and the eta. In a first approximation, these 
GB couple to the spontaneously
broken currents with strength $F\sim 90$ MeV.
These particles are so light compared
with the typical hadronic scales, that they will dominate the
hadronic dynamics at low energies or temperatures.

In order to describe the hadronic dynamics at low energies
we can therefore use these fields to 
build an effective Lagrangian, made of the
most general terms that respect the above symmetry breaking pattern.
As we are interested in the low energy regime, 
the terms are organized according
to their number of derivatives. 
It can be seen, by counting
the powers of momenta of different diagrams, that it is
possible to renormalize any calculation and obtain finite results
order by order in the expansion  \cite{Wei}. 
We could also couple gauge fields, scalar
and pseudoscalar sources, etc..., which would allow us to describe
other processes. This whole approach is
usually known as Chiral Perturbation Theory (ChPT) \cite{GaLe}.

\subsubsection*{Explicit Chiral Symmetry Breaking}

Up to the moment we have just considered the chiral limit.
 When quark masses are turned on,
the GB become massive pseudo-GB 
and their masses can be obtained, generically, as
\bea
M^2_{\pi}&\simeq& 2 B_0 {\hat m} + \Od(m_q^2)\nn\\
M_K^2&\simeq&(\hat{m}+m_s)B_0+  \Od(m_q^2)\nn\\
M_\eta^2&\simeq&\frac{2}{3}(\hat{m}+2m_s)B_0+  \Od(m_q^2)
\label{masaycondensado}
\eea
where $\hat m=(m_u+m_d)/2$ (we will consider 
isospin as an exact symmetry) and 
$B_0$ and other coefficients
that may appear at higher orders are to be determined 
phenomenologically. Throughout this work, the first one 
will play a very relevant role, since it has a very 
physical meaning: In the chiral limit, and up to a 
normalization factor, it is nothing
but the chiral condensate; namely
\be
\qqo\equiv
\langle0\vert \bar u u + \bar d d \vert 0\rangle 
\stackrel{\hat m\rightarrow0}{\longrightarrow} - 2 F_0^2 B_0
\label{chico}
\ee

At this point two different approaches appear in the literature.
The first one, still called ChPT \cite{GaLe}, 
is to assume that the mass expansions 
in eq.(\ref{masaycondensado}) are dominated by the $B_0$ term. 
Its origin can be traced to the Gell-Mann-Okubo (GMO)
 and Gell-Mann-Oakes-Renner (GOR)
formulae, which, within the effective formalism,
 are obtained at first order by eliminating $B_0$ in
eqs.(\ref{masaycondensado}) and (\ref{chico}).
This large condensate scenario usually requires 
$B_0\sim \Od(1\mbox{GeV})$
and, apart from the GMO and GOR formulae, it is supported
by several lattice calculations \cite{lattice}. Within this framework, the
quark masses count as $\Opd$.
The second approach, know as Generalized Chiral Perturbation Theory 
(GChPT)
\cite{GChPT1,GChPT2}, 
is nothing but considering the possibility
that the $\Od(m_q^2)$ terms could be of comparable 
size
or even larger than the $B_0$ term. As a consequence, both the quark
masses and $B_0$ count as $\Od(p)$. This approach is supported by
some deviations from the Goldberger-Treiman relation in $\pi N$, 
$K\Lambda$
and $K\Sigma$ \cite{deviationsGT} and some calculations 
using variationally
improved perturbation theory or a relativistic many body approach
\cite{variationally}.

Those two alternatives are usually compared with the spontaneous 
magnetization $\vec M$ of 
spin systems: On the one hand, ferromagnets present an ordered ground 
state where the magnetization spontaneously acquires an $\vec M \neq 0$ 
value. That would be analogous
to the standard ChPT. On the other hand, in anti-ferromagnets 
the magnetization remains at $\vec M = 0$, which
would be similar to the extreme case of GChPT where $B_0=0$.
Note that, despite their difference, 
in both systems the spins are oriented
in one preferred spatial direction and therefore the  $SO(3)$ rotational
symmetry is broken.

Back to our subject, it should be noticed 
that both approaches have the same terms in the 
Lagrangian, although they are organized differently, and their relative
size is also changed. Indeed, it is possible to reobtain standard ChPT 
as a special case of GChPT.

At present, the experimental data does not exclude 
any of the two scenarios,
although this question may be solved in a few years with an accurate 
measurement of $\pi\pi$
scattering lengths from the decay of $\pi^+\pi^-$ atoms \cite{DIRAC}.

Thus, since we are interested in high temperature differences 
with the standard scenario,
throughout this paper we will use the GChPT formalism.
As usual, the pseudo-GB fields are grouped in an SU(3) matrix
as follows:
\be
U=\exp(i\Phi/F)\quad ; \quad
\Phi=\sqrt{2}\left(
\begin{array}{ccc}
\frac{1}{\sqrt{2}}\pi^0+\frac{1}{\sqrt{6}}\eta&\pi^+&K^+\\
\pi^-&-\frac{1}{\sqrt{2}}\pi^0+\frac{1}{\sqrt{6}}\eta&K^0\\
K^-&\bar{K}^0&-\frac{2}{\sqrt{6}}\eta\\
\end{array}
\right)
\ee 
And then, with the GChPT power counting described above, 
the $\Opd$ Lagrangian is usually written as
\bea
\tilde{\cal L}^{(2)}&=&\frac{4}{F^2}\left\{\tr(D_\mu UD^\mu U^\dagger)+
2B_0\tr({\cal M}(U^\dagger+ U))\right.\nn\\
&+&A_0\,\tr({\cal M}U^\dagger{\cal M}U^\dagger+{\cal M}U{\cal M}U)
+Z_0^S\,\tr({\cal M}(U+U^\dagger))^2\nn\\
&+&
\left. Z_0^P\,\tr({\cal M}(U-U^\dagger))^2 + 2H_0\tr({\cal M}^2)\right\}
\label{GChPTlag}
\eea
where ${\cal M}=diag(\hat m,\hat m, m_s)$ is the quark mass matrix.
In standard ChPT, only the two first terms are $\Opd$, whereas the
rest is counted as $\Opc$.
From the above Lagrangian we obtain
the following meson masses
\bea
M_\pi^2&=&2\hat{m}B_0+4\hat{m}^2 A_0+4\hat{m}(2\hat{m}+m_s)Z_0^S\nn\\
M_K^2&=&(\hat{m}+m_s)B_0+(\hat{m}+m_s)^2A_0+2(\hat{m}+m_s)
(2\hat{m}+m_s)Z^S_0\nn\\
M_\eta^2&=&\frac{2}{3}(\hat{m}+2m_s)B_0+
\frac{4}{3}(\hat{m}^2+2m_s^2)A_0\nn\\
&+&\frac{4}{3}(\hat{m}+2m_s)(2\hat{m}+m_s)Z^S_0+
\frac{8}{3}(m_s-\hat{m})^2Z^P_0
\label{g_masses}
\eea
Comparing with eq.(\ref{masaycondensado}), 
we have just added the $\Od(m_q^2)$
terms. In the standard formalism, since
 only $B_0$ is present, it can be eliminated and one
recovers, at $\Opd$, the GMO and GOR relations. 
That is no longer possible in GChPT, although these relations 
will be recovered at higher orders.
Of the three $\Od(m^2_q)$ parameters 
 there
are two, $Z^0_S$ and $Z^0_P$, which violate 
the Zweig rule. They are expected to be small from
large $N_c$ arguments and is usual to neglect 
their contribution, and so we will do likewise 
in most of what follows.

Note that, since the pion, kaon and eta mass values are known, 
then, changing
the value of $B_0$ is nothing but changing the values of the quark
masses. As a matter of fact, the ChPT relations are
frequently used in the literature to obtain ratios of light quark
masses (for a recent update, see \cite{Leut} and references therein) 
or even to evaluate $\hat{m}$ itself. 
However,  most of these works have used the standard ChPT formalism
and have the large condensate assumption built in, so that their results
would change if it was removed. Nevertheless, there
are determinations of $m_s-m_u$,
which do not rely on a large condensate value.
For the sake of simplicity, and in order to facilitate 
the comparison with previous works \cite{GChPT2}, 
we will use the value $m_s-m_u=(184\pm32)$ MeV,
given in \cite{m_y_r}. That is,
\be
\hat{m}=\frac{184\pm32}{r-1} \mbox{MeV}
\label{m_y_r}
\ee
(There are other similar analyses in ref.\cite{massdif},
whose results are all consistent with the previous relation.)
As a consequence,  the parameter that determines the relative
size of the $\Od(m_q)$ and $\Od(m_q^2)$ terms is the quark
mass ratio $r=m_s/\hat{m}$, which ranges in the interval 
\be
r_1\equiv 2\frac{M_K}{M_\pi}
-1\leq r \leq 2\frac{M_K^2}{M_\pi^2}-1 \equiv r_2
\label{r_tree_limits}
\ee
The upper limit corresponds to the extreme case of a 
very large $B_0$ condensate,
whereas the second corresponds to $B_0=0$. 
(Vacuum stability requires $B_0, 
A_0, Z_0^S\ge0$).

Of course, all these formulae are valid up to $\Opd$. 
For the moment, we have restricted
ourselves to the $\Opd$ case since it already
displays the features of GChPT which are relevant for this work. In
section four we will state our results
including higher order corrections, although we will
just present the GChPT formulae without such 
a detailed introduction.

\subsubsection*{The chiral condensate at zero temperature}

Using the GChPT Lagrangian in eq.(\ref{GChPTlag}), the
chiral condensate at $\Opd$ is then given by  
\be
\qqo=-2F_0^2(B+\hat{m}(A_0+H_0)+...)
\ee
where $B=B_0+2(m_s+2\hat{m})Z^S_0$. In practice $B_0$
 cannot be separated from 
$B$ by looking at quark masses alone, but we have already
commented that $Z_0^S$ is expected to be very small,
so that $B\sim B_0$.
The parameter $H_0$ is associated with the contact term
of two scalar sources, which does not contain meson
fields. However, it is needed as a short distance counterterm, and it
indeed depends on the renormalization conventions, which introduce
some small ambiguity  (see \cite{GaLe} for a discussion).
Nevertheless, using QCD sum rules with a simple model
for the spectral function, and keeping 
$F_\pi^2M_\pi^2$ fixed at its physical value, 
it has been found \cite{Sterncond}  that the chiral condensate
can be described by
\be
\hat{m}\qqo\simeq -F_\pi^2\left[M_\pi^2-4\hat{m}^2 \Omega
\right]
\label{Gcond}
\ee
with $\Omega=4.7\pm0.7$.
At $\Opd$, the $\Omega$ parameter is nothing
but $(A_0-H_0)/2$.
We will use the above equation to estimate the size of the quark
condensate at $T=0$.

In Fig.1 we show the dependence of the 
condensate with $\hat{m}$, for $\Omega=4.7$.
Note that 
the plot starts at $\hat{m}=7$ MeV, which 
is approximately the standard ChPT value.
There,
$\qqo\simeq -(280)^3 \hbox{MeV}^3$, and it decreases
smoothly 
as $\hat{m}$ gets larger, until it vanishes 
around $\hat{m}\simeq 30 \, \mbox{MeV}$. 
The shaded area between dashed lines
covers the uncertainty in $\Omega$.

\begin{figure}
\hspace*{3.5cm}\hbox{
\psfig{file=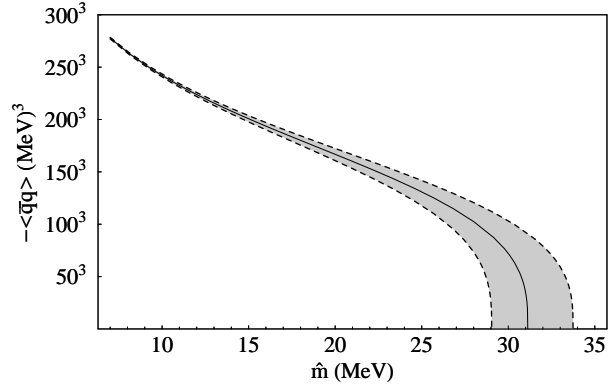,width=8.cm}
}
\caption{ The chiral condensate at zero
temperature as a function of the light quark mass $\hat{m}$,
eq.(\ref{Gcond}). The continuous line corresponds to the central
value of $\Omega$. The uncertainty due to
its error  is covered by the shaded area.}
\end{figure}

As a check of eq.(\ref{Gcond})
we can see that it is consistent with previous estimates
within the framework of standard ChPT \cite{GaLe}, where
\be
\hat{m}\qqo=-F_0^2M_\pi^2
\left[ 1+\frac{M_\pi^2}{32\pi^2F_\pi^2}(4\bar{h}_1+\bar{l}_3-1)\right]
\equiv -F_0^2M_\pi^2 \frac{1}{c}
\ee
and the $\bar{h}_1$ and $\bar{l}_3$  parameters play a similar
role as that of the GChPT $\Opd$ parameters, although in Standard ChPT
they appear at $\Opc$.
In that case, estimates based on a simple $\rho$ resonance model
and the large $N_c$ limit yield $c=0.87$ and $c=0.90\pm0.05$, 
respectively \cite{GeLe}.
If we introduce in eq.(\ref{Gcond}) a value in the range
$\hat{m}$ from 5 to 10 MeV, and take into
account the fact that in standard ChPT $F_\pi/F_0\simeq1.057$, we obtain
$c=0.95\pm 0.4$, which is a highly non-trivial 
check of eq.(\ref{Gcond}). (Throughout this section we have neglected
higher order logarithmic contributions that would yield
corrections of the order of 1\%)

\section{The virial expansion and temperature effects.}

At low energies the free energy $z$ is dominated 
by the contributions from the lightest particles.
Therefore, we can use the Euclidean form of the 
above Lagrangian (denoted $L(x)$) within the standard 
finite temperature functional Euclidean formalism. Hence, 
 in the thermodynamic limit,
\be
z=-T\,\mbox{lim}_{L\rightarrow \infty} \frac{1}{L^3}
\int [dU] \exp\left( -\int_{L^3\times [0,T]}d^4x\, L(x) \right) 
\label{thermal}
\ee
where, as usual, the functional integration is over
pion fields which are periodic in the Euclidean time, 
with period $\beta=1/T$
(see ref.\cite{GeLe}). From the free energy we can derive any other
thermodynamic property of our system, but let us first notice that
since there is a spontaneously broken symmetry, even at $T=0$ there is 
some non-vanishing
vacuum energy density $\epsilon_0$. As a consequence, the pressure
is defined only from the temperature dependent part of the free energy, 
$P\equiv \epsilon_0-z$.

The quark condensate is now obtained as the derivative of
the free energy with respect to the quark mass. That is
\be
\langle \bar q q\rangle\equiv \frac{\partial z}{\partial \hat m}=
\langle 0 \vert \bar q q\vert 0 \rangle-
\frac{\partial P}{\partial\hat m}
\label{qq_and_P}
\ee
where we have used that
at $T=0$ the condensate is nothing but the vacuum expectation value
$\langle 0 \vert \bar q q\vert 0\rangle
\equiv \partial \epsilon_0/\partial 
\hat m$.

In this section we will
just concentrate on how to obtain
$\partial P/\partial \mh$.
For that purpose, one possibility is to calculate the free energy 
from the effective Lagrangian, as was done in \cite{GeLe} 
within standard ChPT. That method follows the very same philosophy
of the chiral expansion, but is rather lengthy.
In this paper we will make use of existing one loop calculations 
of elastic $\pi\pi$ scattering, together with the relativistic 
virial expansion
of a pion gas \cite{Dashen,nosotros}. 

Let us then consider a gas made only of pions. This approximation seems 
reasonable as long as the temperatures remain sufficiently 
below the kaon threshold
\cite{GeLe}.
Within the virial formalism, the pressure can be expanded  as follows

\begin{equation}
P=3T\left(\frac{M_{\pi}T}{2\pi}\right)^{3/2}\sum_{k=1}^{\infty}B_k
e^{-\beta M_{\pi}k}=
3\frac{T}{\lambda^3}\sum^{\infty}_{k=1}B_k(T) \xi^k.
\label{pvir}
\end{equation}
The factors of three come from the fact that we are neglecting isospin
breaking effects. Thus, there are effectively 
three different species of
particles, labeled according to their third isospin component, 
that behave identically with respect to strong forces.
We have also defined $\lambda=(2\pi/M_{\pi}T)^{1/2}$, 
which  is the thermal 
pion wavelength. Note that the expansion parameter is the 
fugacity $\xi=e^{-\beta M_{\pi}}$. In a non-relativistic framework, 
the expansion is usually performed using the definition 
$\xi=e^{\beta \mu}$,
where $\mu$ would be the chemical potential. In contrast, 
in the relativistic
case, there is a rest energy given by $M_\pi$, whose contribution
to eq.(\ref{thermal}) 
is equivalent to a chemical potential $\mu=-M_\pi$ in a non-relativistic
description.

There is a closed expression for the  
virial coefficients of the free gas, which is
\begin{eqnarray}
B_n^{(0)}(T)&=&\frac{3}{n(M_{\pi}T)^{3/2}}\sqrt{\frac{2}{\pi}}
\int^{\infty}_{0}dp\;p^2
 e^{-n\beta (E(p)-M_{\pi})}.
\label{virfree}       
\end{eqnarray}
In order to deal with the interacting gas, we will just consider
two particle interactions, which can be justified as long as the 
density remains small. In \cite{GeLe} it was shown that this
is consistent with the three loop calculation in ChPT.
In such case, it is enough to keep the two first
terms of the virial expansion, whose coefficients will be 
given by \cite{Dashen}
\bea
B_1(T)&=&B_1^{(0)}(T)\label{Bint}\\ \nonumber
B_2(T)&=&B_2^{(0)}(T)+\frac{4e^{2M_{\pi}/T}}{(2\pi M_{\pi}T)^{3/2}}
\int^{\infty}_{2M_{\pi}}
dE E^2 K_1(E/T)\left(\sum_{I,J}(2I+1)(2J+1)\delta_{IJ}(E)\right)
\eea
where $K_1(x)$ is the modified Bessel function which behaves as 
$\sqrt{\pi/2x}e^{-x}$
when $x\rightarrow\infty$. It is important to notice that the  only 
dynamical information we need are the phase shifts $\delta_{IJ}$.
As an estimate of the applicability of this approach, it was shown in
\cite{nosotros} that the second order virial expansion yields
less than a 1\%
error when applied to the free gas up to $T\sim 250$ MeV.

Let us finally remark that the high temperature 
behavior of the chiral condensate will be due to
two different effects. First, the starting $T=0$ value, which
may differ from the standard, large condensate, value. But, second, it
also depends on how the mass dependence of the phase shifts has changed
with respect to standard ChPT.

In the next section the phase shifts
will we obtained using the existing GChPT calculations of the
$\pi\pi$ elastic scattering amplitudes \cite{GChPT2}.
 In later sections we will include 
contributions from particles more massive than pions.

\section{The General Scenario}

\subsubsection*{Higher orders in GChPT}

Within the standard ChPT it was shown in \cite{GeLe} that 
the $\Opc$ contributions accelerate the melting
of the chiral condensate, lowering the critical temperature. 
Our aim now is to include the equivalent corrections within GChPT.
Unfortunately, we have already seen that the $\tilde{\cal L}^{(2)}$ 
Lagrangian
has more terms that the standard ${\cal L}^{(2)}$. That means that
there are many more phenomenological parameters in the Lagrangian,
which in many cases are not very well known.
The situation gets
even worse at higher orders. In general, the GChPT Lagrangian is
built of terms like \cite{GChPT1,GChPT2}
\be
\tilde{\cal L}^{(d)}=\sum_{k+l+n}B_0^n{\cal L}_{(k,l)}, 
\quad \mbox{with}\quad
{\cal L}_{(k,l)}\sim \Od(p^k m_q^l)
\ee
Indeed, we have already given 
$\tilde{\cal L}^{(2)}$ in eq.(\ref{GChPTlag}) 
and we found that  some of the constants are not very well determined.
For the complete expression of the 
 $\Opc$ Lagrangian we refer to \cite{GChPT2}. 
For our purposes, there are several relevant modifications 
to our previous discussion: First, 
now there are three different decay constants $F_\pi$,
$F_K$ and $F_\eta$. Second, neglecting Zweig rule 
violating parameters, the expressions for $M_\pi$ and $M_K$ in 
eq.(\ref{g_masses}) 
are now modified to
\bea
\frac{F^2_\pi}{F^2}M_\pi^2\
&=&2\hat{m}B_0+4\hat{m}^2 A_0+
\frac{F^2_\pi}{F^2}\delta M_\pi^2\nn\\
\frac{F^2_K}{F^2}M_K^2
&=&(\hat{m}+m_s)B_0+(\hat{m}+m_s)^2A_0+\frac{F^2_K}{F^2}\delta M_K^2
\label{g_loop_masses}
\eea
where $\delta M^2_i$ are higher order corrections 
and logarithmic terms \cite{GChPT2}, 
whose size is $\delta M^2_i<0.1 M^2_i$ \cite{SternMainz} (see below).
As a consequence, the range of allowed $r$ values is shifted upwards to
\be
r^*_1\equiv 2\frac{F_KM_K}{F_\pi M_\pi}
-1\leq r \leq 2\frac{(F_K M_K)^2}{(F_\pi M_\pi)^2}-1 \equiv r^*_2.
\label{r_loop_limits}
\ee
With this modifications now
$r^*_1\simeq 8$ and $r_2^*$ can be as large as 39.

Finally, there are also higher order corrections to the $T=0$ condensate
itself, which contain chiral logarithms. That means that we 
cannot simply say that  $\Omega=(A_0+H_0)/2$. Nevertheless we can still
use the phenomenological parameter $\Omega=4.7\pm0.7$ in
eq.(\ref{Gcond}).

\subsubsection*{The one loop $\pi\pi$ amplitude in GChPT}

Next, we need the 
$\pi^+\pi^-\rightarrow\pi^0\pi^0$ scattering 
amplitude itself. Although it has been calculated
in GChPT up to two loops \cite{GChPT1,GChPT2},
for our purposes it will
be more than enough to consider the one loop result,
which reads:
\bea
A(s,t,u)&=&\frac{\alpha}{3F_\pi^2}M_\pi^2+
\frac{\beta}{F_\pi^2}\left(s-\frac{4}{3}M_\pi^2\right)\label{pipi}\\
&&+\frac{\lambda_1}{F_\pi^4}(s-2M_\pi^2)^2+
\frac{\lambda_1}{F_\pi^4}\left[(t-2M_\pi^2)^2+(u-2M_\pi^2)^2\right]
+\bar{J}_{(\alpha,\beta)}(s,t,u) \nn
\eea
where
\bea
\bar{J}_{(\alpha,\beta)}(s,t,u)&=&
\frac{1}{6F_\pi^4}\left\{
4\left[\frac{5}{6}\alpha M_\pi^2+\beta\left(s-\frac{4}{3}M^2_\pi
\right)\right]^2-
\left[\frac{2}{3}\alpha M_\pi^2-\beta\left(s-\frac{4}{3}M^2_\pi
\right)\right]^2 \right\}\bar{J}(s)\nn\\
&&+\frac{1}{12F_\pi^4}\left\{
3\left[\frac{2}{3}\alpha M_\pi^2-\beta\left(t-\frac{4}{3}M^2_\pi
\right)\right]^2+\beta^2(s-u)(t-4M_\pi^2)
 \right\}\bar{J}(t)\nn\\
&&+\frac{1}{12F_\pi^4}\left\{
3\left[\frac{2}{3}\alpha M_\pi^2+\beta\left(u-\frac{4}{3}M^2_\pi
\right)\right]^2+\beta^2(s-t)(u-4M_\pi^2)
 \right\}\bar{J}(u)
\eea
and $\bar{J}$ is the standard one-loop integral \cite{GaLe}.

In the literature, the 
values of the $\alpha$, $\beta$, $\lambda_1$ and $\lambda_2$
are fitted from experiment. However, in order to obtain the 
condensate dependence with the temperature, we
need the derivative of the pressure with respect to $M_\pi$,
and just a fitted value is not enough. Therefore, we also need to know
the $M_\pi$ dependence of the parameters, and, if we want
to study the effects of changing the light quark masses, we also
need the dependence on $r$.

\subsubsection*{Phenomenological parameters}

The actual expressions of the $\alpha$, $\beta$,
parameters are rather complicated and involve many parameters from
the GChPT Lagrangian, which frequently are not very well determined.
In addition they contain chiral logarithms. It is therefore
very convenient to
expand $\alpha$ and $\beta$ in powers of quark masses, namely
\be
\alpha=\sum_{n=0}^3 \alpha^{(n)},\quad \beta=\sum_{n=0}^3 \beta^{(n)},
\ee
Notice that, in GChPT, since quark masses
are considered as $\Od(p)$, these expansions involve
terms that count as {\em odd} powers of momenta.

The above expansions have been worked out in \cite{GChPT2}, and they
are the following:
\bea
\alpha(r)&=&1+6\frac{r_2^*-r}{r^2-1}
-\frac{4}{r-1}\left(\frac{F_K^2}{F^2_\pi}-1\right)
+\,18(2-r)\hat\rho_1
-\,6\,r\,\hat\rho_2+\alpha^{(2)}(r)\nn\\
\beta(r)&=&1+\frac{2}{r-1}\left(\frac{F_K^2}{F^2_\pi}-1\right)
+\beta^{(2)}(r)
\label{ab_evo}
\eea
where in all the above equations we have neglected the Zweig rule
violating parameters.

Let us now try to estimate the size of the different terms
in the $\alpha$ and $\beta$ expansions. Let us then look back
at the allowed values of $r$, eq.(\ref{r_loop_limits}).  The relevant
point for our discussion is that now, even with the lowest value $r=8$,
we obtain, using eq.(\ref{m_y_r}), that  $\mh\leq (26\pm4.6)$ MeV. 
Therefore we can estimate that the terms coming from $B_0$ and $A_0$ 
should be $\Od(1)$, those from $\tilde{\cal L}^{(3)}$ should be $\Od(10\%)$
and those from ${\cal L}_{(2,2)}$ and ${\cal L}_{(0,4)}$ should at most
reach the 1\% level.
Consequently, we will neglect the $\alpha^{(2)}$ and $\beta^{(2)}$ effects.
The only parameters that remain undetermined are the $\hat\rho_{1,2}$, 
which contribute to $\alpha^{(1)}$. However, from a dimensional analysis
\cite{GChPT2},
their magnitude can be naively estimated as 
$\vert \hat \rho_i\vert\simeq (0.4\pm0.2)/(r-1)^3$.
 Their dependence on the actual
value of $M_\pi$ (needed for the numerical derivation) seems very 
weak.
In our calculations we will 
take them first as zero and then we will include them in the 
uncertainty.

Concerning $\lambda_1$ and $\lambda_2$, they come only
from the terms in ${\cal L}_{(4,0)}$, 
which do not contain explicit chiral
symmetry breaking. They are given by
\bea
\lambda_1&=&\lambda_1^{(0)}=4(2L_1^r(\mu)+L_3)-\frac{1}{48\pi^2}\left\{
\log\frac{M^2_\pi}{\mu^2}+\frac{1}{8}\log\frac{M^2_K}{\mu^2}
+\frac{35}{24}\right\}\nn\\
\lambda_2&=&\lambda_2^{(0)}=4L_2^r(\mu)-\frac{1}{48\pi^2}\left\{
\log\frac{M^2_\pi}{\mu^2}+\frac{1}{8}\log\frac{M^2_K}{\mu^2}
+\frac{23}{24}\right\}
\label{l_evo}
\eea
It can be seen that these parameters do not carry any $r$
dependence. For definiteness, we will use for them the values obtained
in \cite{GChPT2}:
\be
\lambda_1=(-5.3\pm2.5)10^{-3};\quad \lambda_2=(9.7\pm1.0)10^{-3},
\ee
which are consistent with other determinations in the standard framework.

The values of $\alpha$ and $\beta$ depend on whether there is actually
a large or small condensate at $T=0$, and we will use their
$r$ dependence to reproduce different scenarios. For illustrative
purposes, let us recall that in the standard formalism both $\alpha$
and $\beta$ are slightly bigger than one and $r\simeq26$.
In contrast, the low condensate alternative seems to prefer
$\alpha\simeq2$ and $r\simeq10$ \cite{GChPT2}.

\subsubsection*{Phase shifts}

For the virial expansion we need the phase shifts of
definite isospin and angular momentum channels. 
At lowest order, they are defined as follows (see \cite{phases} for
a discussion on this subject)
\be
\tan \delta_{IJ}(s)= \sigma(s) \Rea( t_{IJ}(s)),
\ee
where $\sigma(s)=\sqrt{1-4M_\pi^2/s}$. The partial waves $t_{I,J}$ are
obtained from the isospin amplitudes
\bea
T_0(s,t,u)&=&3A(s,t,u)+A(t,s,u)+A(u,t,s),\nn\\
T_1(s,t,u)&=&A(t,s,u)-A(u,t,s),\nn\\
T_2(s,t,u)&=&A(t,s,u)+A(u,t,s),
\eea
by means of
\be
t_{IJ}= \frac{1}{64\pi}\int^1_{-1}d(\cos\theta)P_J(\cos\theta)T_I(s,t,u).
\ee
where $P_I$ is the corresponding Legendre polynomial.
In our calculations we have just used the lowest angular momentum
for each isospin channel, namely $(I,J)=(0,0), (1,1)$ and $(2,0)$.
For all means and purposes, they 
dominate the low energy pion interactions.

\subsubsection*{The calculation of $\partial P/\partial \mh$}

We have then used
the above phase shifts with the second order
virial expansion. In order to
obtain the condensate,  eq.(\ref{qq_and_P}), we then need 
$\partial P/\partial \mh$, which can be obtained using
\be
\frac{\partial P}{\partial \mh}=
 \frac{\partial P}{\partial M_\pi}\,\frac{\partial M_\pi}{\partial \mh}+
 \frac{\partial P}{\partial M_K}\,\frac{\partial M_K}{\partial \mh}+
 \frac{\partial P}{\partial F_\pi}\,\frac{\partial F_\pi}{\partial \mh}+
 \frac{\partial P}{\partial F_K}\,\frac{\partial F_K}{\partial \mh}
\label{P_derivatives}
\ee
Naively one just expects the first term, but let us remember that $M_K$,
$F_\pi$ and $F_K$ are $\mh$-dependent and they appear in the
amplitude either
directly or indirectly through $\alpha$, $\beta$, $\lambda_1$ and $\lambda_2$.
That problem was carefully avoided in \cite{GeLe} by using $SU(2)$ 
standard ChPT and only using $F$ in the free energy expansion.

Of course, only $M_\pi$ appears in the fugacity, or in the free gas virial
coefficients and thus we expect the three last terms in 
eq.(\ref{P_derivatives}) to be much smaller than the first. 
Indeed, within the range of $r$ and $T$ that we are interested in,
we have found that the term due to the appearance
of $M_K$ in the amplitude is smaller than $1\%$ and we have neglected it.
In contrast, $F_\pi$ and $F_K$ together generate contributions of
the order of $5\%$, and therefore they have been included in our calculations.

The derivative of the pressure with respect to $M_\pi$, $F_\pi$ and $F_K$
have been performed numerically, with an increment of $0.1$MeV.
For instance, the value of the pressure is first
calculated with the real $M_\pi$ and then with $M_\pi-\delta M_\pi$,
{\em including a change in  the chiral parameters}, following
eqs.(\ref{ab_evo}) and (\ref{l_evo}). A similar procedure is followed
for $F_\pi$ and $F_K$.

In our calculations we have used
\bea
\frac{\partial M_\pi}{\partial\hat m}&\simeq&
\frac{M_\pi}{2\mh}\left[1+2\frac{r_2^*-r}{r^2-1}\right]\nn\\
\frac{\partial M_K}{\partial\mh}&\simeq&
\frac{M_\pi^2}{4\mh M_K}\frac{r(2r_2^*-r)-1}{r^2-1}\nn\\
\frac{\partial F_\pi}{\partial \mh}&\simeq& 
\frac{F_\pi}{\mh \left[ (r-1)+ \left(F_K^2/F_\pi^2-1\right)\right]}
\left(\frac{F_K^2}{F_\pi^2}-1\right)\nn\\
\frac{\partial F_\pi}{\partial \mh}&\simeq&
\frac{r-1}{2} \frac{\partial F_\pi}{\partial \mh}
\eea
which we have obtained from eqs.(\ref{g_loop_masses})
and from \cite{GChPT2}. There are, of course, corrections,
but their effects on the final results are again less than $1\%$.

\subsubsection*{Results}

As we have already commented, the virial expansion can be trusted 
only at low temperatures, mostly, due to the fact that 
above $\sim 150$ MeV the contributions from other 
more massive particles becomes relevant. These effects will be studied
in the next section and we will see that they tend to
lower the critical temperature, which is therefore
more favorable for our approach. For the moment, 
if we give in our figures results for higher temperatures, they should
be interpreted with great care as a qualitative behavior or, for
instance, as a tendency towards symmetry restoration.
Nevertheless, 
comparing between different figures could
also illustrate what is the qualitative effect of a change in the
parameters.

Thus, in Fig.2 we have shown the
dependence of the chiral condensate with the temperature
and for a light quark ratio in the range $8\leq r\leq 26$. 
For $\Omega$ we have used the central value 4.7.
Although the actual points
at which $\qq=0$ are just gross estimates,
we can see that lowering $r$ yields a systematic decrease
in the chiral phase transition temperature.

\begin{figure}
\hspace*{3cm}
\hbox{
\psfig{file=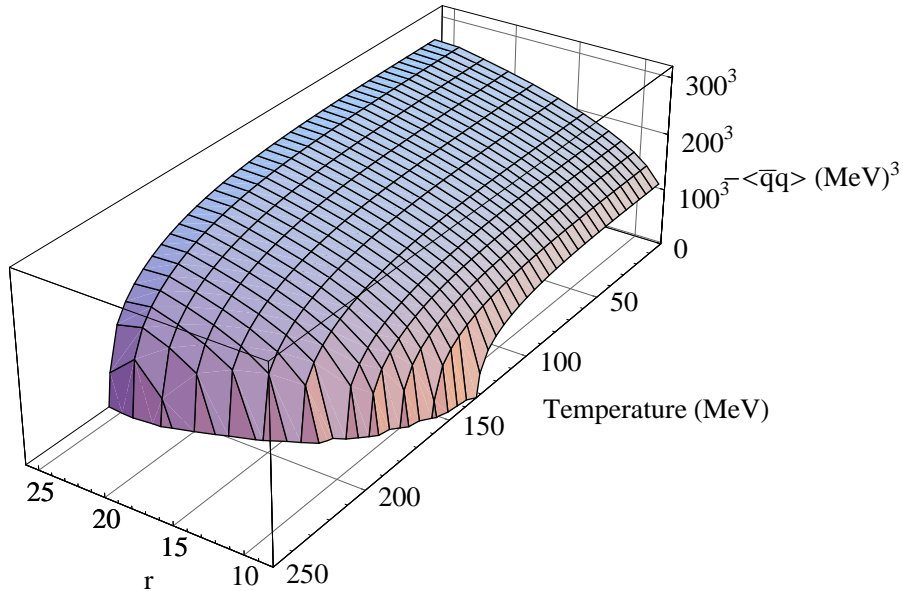,width=12.cm}
}

\vskip 1.cm

\caption{ Quark condensate versus 
temperature and $r$ for $\Omega=4.7$.}
\end{figure}

Indeed, $T_C$ seems to be above 200 MeV for $r\geq 20$ going down to
 130 MeV around $r=8$. Note that for the latter
temperature our approximations can become quite reliable. 
As we have already 
seen, smaller values of $r$ are forbidden to ensure vacuum stability.

The previous results have been obtained using the central values of 
all parameters. In Fig.3a we show what happens if we take into account
the uncertainty in $\Omega$ and  thus, we plot 
the temperature dependence
for the two extreme cases, $r=26$ and $r=8$. The former, which
corresponds to the upper curve, is almost insensitive to this variation.
It corresponds to the standard formalism, where the value of
the chiral condensate is largely 
dominated by the $\Od(m)$ term and, consistently, 
changes in the other terms are almost negligible.
On the other curve, which is associated with the lowest condensate 
scenario, the effect of this error is translated to a 10 to 15\% 
change in $T_C$, at most. 

\begin{figure}

\hbox{
\psfig{file=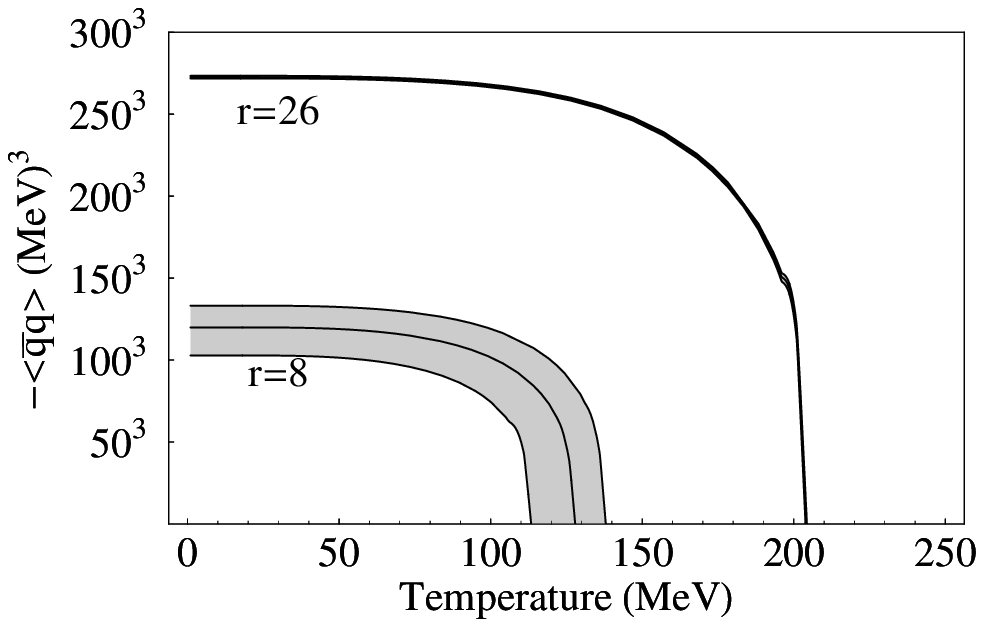,width=7.5cm}
\psfig{file=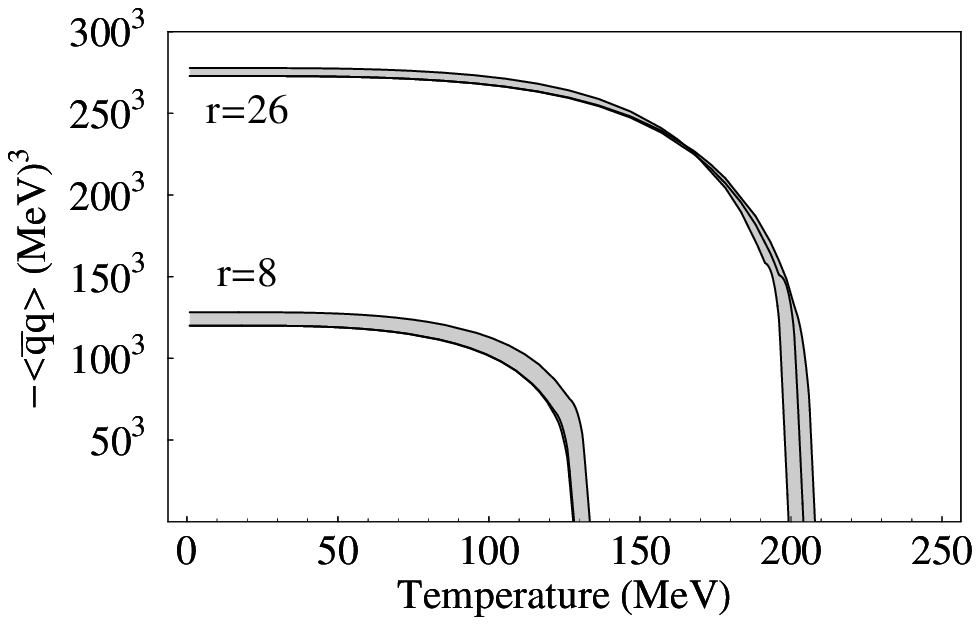,width=7.5cm}
}

\caption{ a) Estimate of the errors in the $\qq$ evolution
due to the uncertainty in  $\Omega$.
b) Uncertainties in the parameters that appear in the
$\pi\pi$ phase shifts, added linearly. }
\end{figure}

In Fig. 3b we show the uncertainties associated with all the parameters that
appear in the scattering amplitude. In the shaded areas, we have 
taken into account  all the effects of changing $\lambda_1$, $\lambda_2$, 
$\hat{\rho}_1$ and $\hat{\rho}_1$ . 
In addition we have also let
the pion and kaon masses vary between their values for the neutral 
or scalar particle. Note that in the case of the pion mass, such a change
also affects the coefficients of the virial expansion and the fugacity.
Finally, we have included the uncertainty
in $F_K/F_\pi=1.22\pm0.01$ and we have let $F_\pi$ change between
92.4 and 93.2 MeV which are two values currently cited in the literature.
Both $M_\pi$ and $F_\pi$ do also appear in the expression of the 
$T=0$ chiral condensate.
All in all, the overall uncertainty in $T_C$ 
due to these parameters seems 
to be of the order of $\pm5$ MeV at $r=26$ and $\pm3$ MeV at $r=8$.
Since we have just simply
{\em added} the different 
errors, we consider these numbers as a conservative
estimate.

\section{Other massive particles}

In this section we will consider the effect of
adding heavier particles to our pion gas. 
We will be following closely the approach of Gerber and Leutwyler
\cite{GeLe} with slight modifications to implement also the
low condensate scenario.

The density of massive states should be exponentially suppressed
by Boltzmann factors $\exp{(-M_i/T)}$, which means that their two
body interactions will carry an $\exp{[-(M_i+M_j)/T]}$ factor.
In addition, their interactions with pions are also suppressed 
by $T^2/F^2$,
due to the chiral symmetry.
Hence, we can treat those heavier particles in the
free gas approximation. In such case, we have an additional contribution
to the pressure, which is given by
\be
\Delta P=-\sum_i\frac{g_i T}{2\pi^2}\int^\infty_0 dp \,p^2 
\log\left[1-e^{-\sqrt{p^2+M_i^2}/T}\right]
\ee
where $g_i$ is the state degeneracy  of a state with mass $M_i$
(that was the
 factor of 3 in eq(\ref{pvir})). 
Note that,
since we will be dealing with temperatures much smaller
than the first hadronic fermions, it makes sense just to use
Bose statistics.
The above formula is only meaningful at low temperatures, since
as we increase the temperature, the mean distance between
massive states shrinks and the dilute gas approximation
is no longer valid. In ref.\cite{GaLe} it was estimated that 
the model is valid up to temperatures
on the order of 150 MeV, although it ``rapidly deteriorates'' 
for higher temperatures. 

Back to the condensate, and in view of eq.(\ref{qq_and_P}), the 
new contributions are of the form
\be
\Delta\qq = -\sum_i
\frac{\partial \Delta P}{\partial M_i}
\frac{\partial M_i}{\partial\hat m}
\label{mass_cont}
\ee
and therefore
\be
\Delta\qq = \frac{1}{2\pi^2}\sum_i g_i\,M_i\,
\frac{\partial M_i}{\partial\hat m}\,
\int_0^\infty dp\, \frac{p^2}{\sqrt{p^2+M_i^2}}
\,\frac{1}{e^{\sqrt{p^2+M^2_i}/T}-1}.
\ee
Thus, we only have to estimate the value of 
$\partial M_i/\partial\hat m$. 
Naively, 
one would expect that the contribution $\hat{m}$ 
to a hadron mass would be roughly proportional to 
the number $N_i$ of $u$ and $d$ quarks it contains.
That estimate seemed quite appropriate
in the standard framework \cite{GeLe}.
We now have to check that it is also the case
in GChPT.

Let us then go back to eqs.(\ref{g_masses}), since
to get rough estimates it is enough to work
at $\Od(p^2)$. As usual, we neglect the $Z_0^S$ and $Z_0^P$
parameters. Then, we obtain the following derivatives
\begin{eqnarray}
\frac{\partial M_K}{\partial\mh}&\simeq&
\frac{M_\pi^2}{4\mh M_K}\frac{r(2r_2-r)-1}{r^2-1}\nn\\
\frac{\partial M_\eta}{\partial\hat m}&\simeq&
\frac{M_\pi^2}{6\mh M_\eta}\left[1+2\frac{r_2-r}{r^2-1}\right]
\label{dMdm}
\end{eqnarray}
We can reproduce the standard scenario with $r=26$, which yields
$\mh\simeq7.4\pm1.3$ MeV using eq.(\ref{m_y_r}). In such case,
we find $\partial M_K/\partial\hat m\simeq1.3\pm0.2$,
which is in very good agreement with a rough estimate of 1.
We also find $\partial M_\eta/\partial\hat m\simeq0.8\pm0.1$,
again consistent with the naive estimate of $2/3$. In any case,
it seems that $\partial M_i/\partial\hat m=N_i$ is a 
small underestimate  of the
actual values of the standard scenario, as it was already pointed
out in \cite{GeLe}. Thus, in that work they
considered that the range from $N_i$
 to $2N_i$ was a ``fair representation'' of the uncertainty in
 $\partial M_i/\partial\hat m$.

However, if we set $r=8$, which corresponds to
 the lowest allowed $T=0$ condensate,
we find $\partial M_K/\partial\hat m\simeq2.0\pm0.4$
and $\partial M_\eta/\partial\hat m\simeq0.36\pm0.06$.
Again, the order of magnitude is correct, although within a factor of 2.
We will therefore use the estimates in 
eq.(\ref{dMdm}) for the kaon and
the eta, instead of $\partial M_i/\partial\hat m=N_i$.
Those are the states that will contribute more at low temperatures.
For the rest, we will assume the uncertainty
in $\partial M_i/\partial\hat m=N_i$ to be from $N_i/2$ to $2N_i$.

\begin{figure}
\hspace*{3cm}
\hbox{
\psfig{file=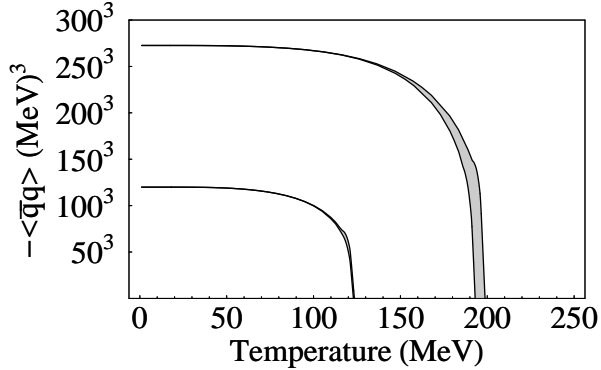,width=8.cm}
}

\caption{The evolution of the chiral condensate when we include
corrections from a free gas of particles more massive than the pion.
The shadowed regions cover the uncertainties in $\partial M_P/\partial \mh$
described in the text. These contributions always tend to lower 
the critical temperature.
}
\end{figure}

Thus, in Fig.4 we show the results when the massive states are taken 
into account. We have considered in eq.(\ref{mass_cont}) all 
particles containing $u$ and $d$ quarks up to $1300 \, \mbox{MeV}$
and we have taken the central values of all the other parameters.
The dominant contributions are, of course, those of the kaons, 
the eta, the rho and the omega. The shaded areas cover the uncertainty 
in $\partial M_i/\partial \mh$ that we have just described. 
Obviously, the net effect is biggest for the
standard scenario, since the critical temperature is higher, where
$T_C$ is decreased down to 190 to 200 MeV. This result, 
although it has been
obtained within the 
generalized formalism, reproduces very nicely the standard ChPT
estimate given in \cite{GeLe}.

\begin{figure}
\hspace*{2cm}
\hbox{
\psfig{file=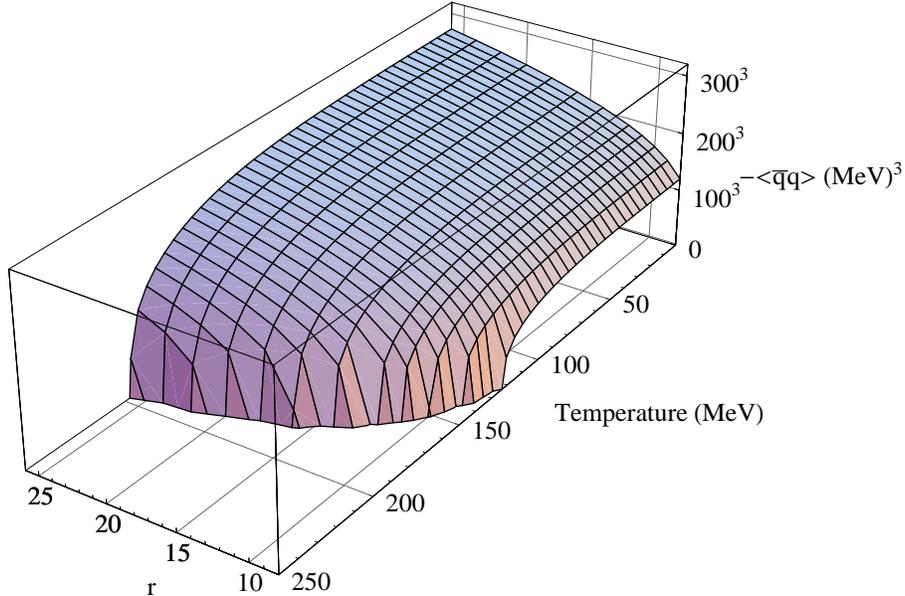,width=12.cm}
}

\vspace*{.5cm}

\caption{The evolution of the chiral condensate when we include
corrections in the pion gas from heavier states, as a function
of the temperature and the quark mass ratio $r$. (Using the central values of
all parameters and estimates)
}
\end{figure}

 Indeed, the $r$
dependence is given in Fig.5 where  we plot
the evolution of the chiral condensate both with $T$ and $r$, for the central
values of all the parameters, but also including the contributions from massive
states. Note that, for the extreme case when $r=8$, the decrease is of the 
order of 5 MeV, down to around 125 MeV.

\section{Conclusions}

In this work we have studied the generalized scenario
of chiral symmetry breaking, either with a large or a small
$T=0$ condensate. For that purpose we have described a pion gas 
by means of the
virial expansion, whose coefficients have been calculated
using the amplitudes obtained within $\Opc$ Generalized Chiral 
Perturbation Theory.

We have also added a crude estimate of contributions from 
particles heavier than the pion, in a free gas approximation, which can be
justified at low temperatures. The effect of these particles is
always to {\em decrease} the temperature of chiral restoration.
Their net effect is to lower $T_C$ by 10 to 20 MeV
in the standard scenario, and by around 5 MeV when the
$T=0$ chiral condensate is smallest.

From our results, it seems that 
the chiral phase transition in a pure pionic gas
may occur at energies as low as 125 MeV in the lowest possible 
$T=0$ condensate scenario. The main source of uncertainty
is the fact that within the Generalized 
approach many parameters still remain undetermined. In the worst case,
which again corresponds to the lowest condensate and lowest $T_C$, it
can be estimated at about 20\%. 
For the standard case of a large condensate, we recover 
previous estimates of $T_C\simeq 190\,\mbox{MeV}$.

In conclusion, we have found that the value of $\sim 190$ MeV 
for the critical 
temperature obtained from standard Chiral Perturbation theory can be
seen as an upper bound if we were to include $\Od(m^2_q)$ corrections
in the mass terms, in addition to the standard condensate contribution.
The effects of these corrections always
lower the critical temperature, which, all together, could be as low 
as $125\,\mbox{MeV}$ with a 20\% 
uncertainty for the lowest condensate scenario.

\section*{Acknowledgments}

I am especially indebted to A.Dobado for introducing 
me to the subject of pion gas thermodynamics, as well as 
J. Stern for explaining me the present status of
some error estimates, D. Espriu for several comments and suggestions
and J. Wells for a careful reading of the manuscript.
I would also like to thank the Theory Group at SLAC for their 
kind hospitality and the Spanish
Ministerio de Educaci\'on y Cultura for a Fellowship. This work 
has been partially supported by the 
Spanish CICYT under contract AEN93-0776.

\end{document}